\documentclass[aps,prb,longbibliography,10pt
]{revtex4-2} 

\usepackage{graphicx,graphics,amsfonts,color,bm,amsmath,hyperref, 
amssymb,wasysym,ulem}
\usepackage[ansinew]{inputenc}

\def\be{\begin{equation}} 
\def\ee{\end{equation}} 
\def\ba{\begin{eqnarray}} 
\def\ea{\end{eqnarray}} 
\def\bc{\begin{center}} 
\def\ec{\end{center}}

\def\p{\partial}

\DeclareMathOperator\sign{sign}

\begin{document} 

\title{Theory of the in-plane photoelectric effect in quasi-one-dimensional electron systems} 

\author{S. A. Mikhailov} 
\email[Electronic mail: ]{sergey.mikhailov@physik.uni-augsburg.de} 
\affiliation{Institute of Physics, University of Augsburg, D-86135 Augsburg, Germany} 

\date{\today} 

\begin{abstract}
The in-plane photoelectric (IPPE) effect is a recently discovered [Sci. Adv. \textbf{8}, eabi8398 (2022)] quantum phenomenon which enables efficient detection of terahertz (THz) radiation in semiconductor structures with a two-dimensional (2D) electron gas. Here we develop a theory of the IPPE effect in quasi-one-dimensional electron systems in which the width of the 2D conducting channel is so small that the transverse quantization energy is larger than the thermal energy. We calculate the THz photoresponse of such a system, as a function of the THz frequency, control gate voltages, and geometrical parameters of the detector. We show that the transverse quantization of the electron motion manifests itself in oscillating gate-voltage dependences of the photocurrent, if the THz photon energy is less than the one-dimensional quantization energy. Results of the theory are applicable to any semiconductor systems with 2D electron gases, including III-V structures, silicon-based field effect transistors, and the novel 2D layered, graphene-related materials.
\end{abstract} 


\maketitle 


\section{Introduction\label{sec:intro}}

\begin{figure}[ht]
\includegraphics[width=8.5cm]{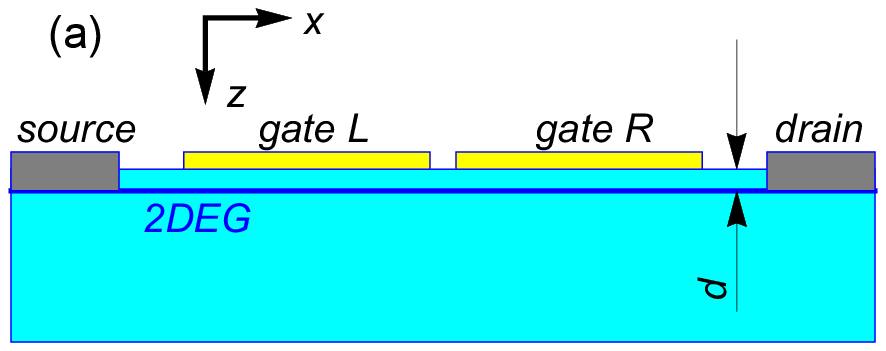}  \\
\vspace{1.5mm}
\includegraphics[width=8.5cm]{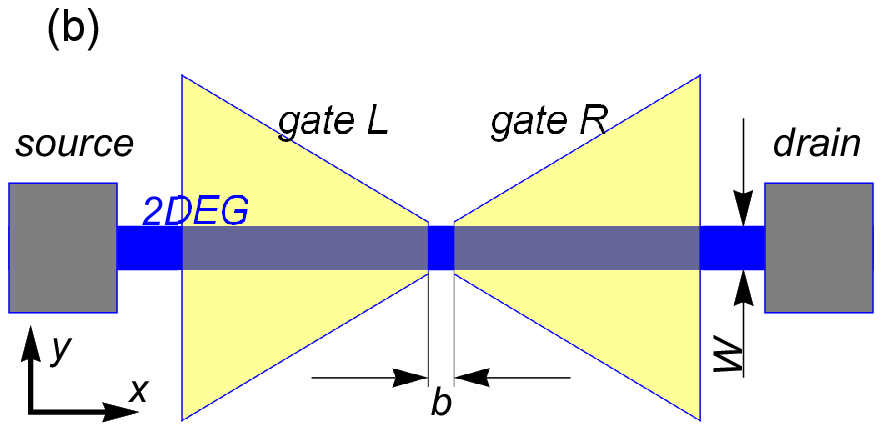}  \\
\vspace{1.5mm}
\includegraphics[width=8.5cm]{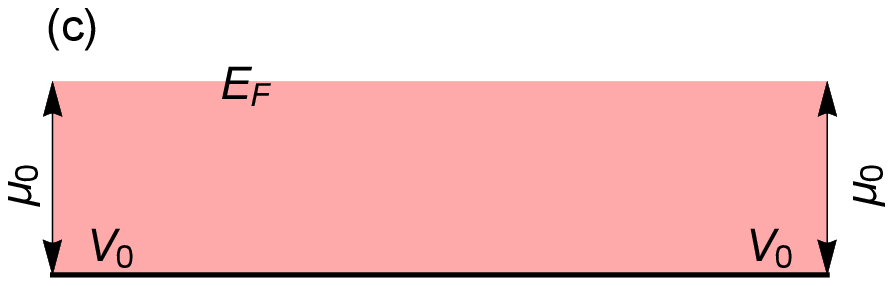} \\
\vspace{1.5mm}
\includegraphics[width=8.5cm]{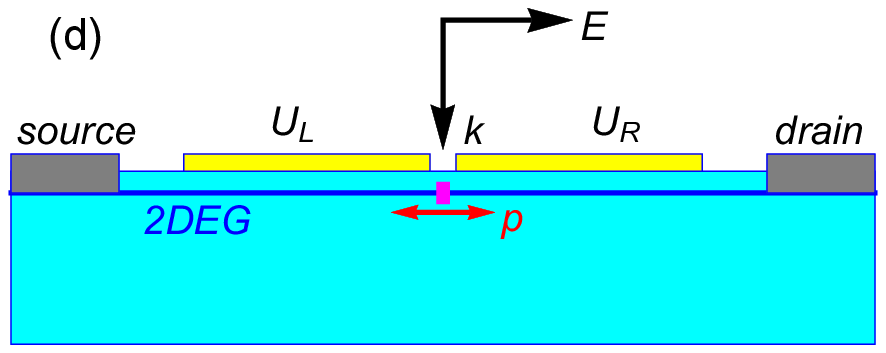} \\
\vspace{1.5mm}
\includegraphics[width=8.5cm]{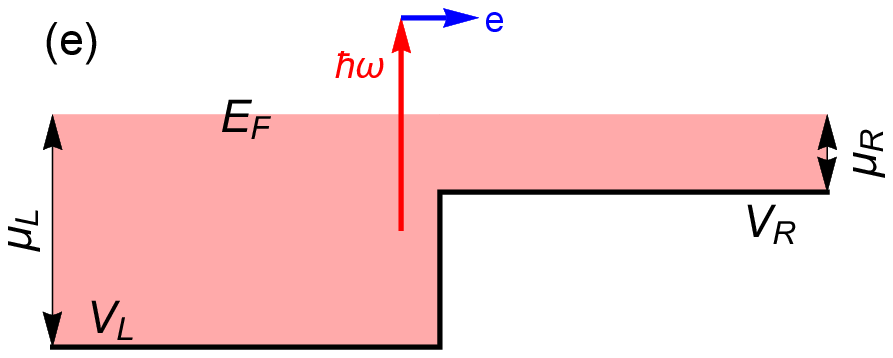} 
\caption{\label{fig:physics} The geometry and the operation principle of the photoelectric tunable-step (PETS) detector based on the IPPE effect: (a) side, and (b) top view of the device; (c) band structure in equilibrium, at $U_{L}=U_{R}=0$; (d) different gate voltages $U_{L}\neq U_{R}$ create a potential step (illustrated by the magenta rectangle) on the line $x=0$ in the lateral ($x$-) direction; under THz irradiation electrons oscillate with a large momentum $\bm p$ (red arrows) in the area under the gap between the gates, absorb THz quanta and jump on the potential step; (e) band structure and the THz photoexcitation process at the potential step. The width of channel $W$ is assumed to be so small in this paper that the transverse quantization energy $E_\perp$ exceeds the temperature $T$, $E_\perp\gtrsim T$ (in Ref. \cite{Mikhailov22} the opposite case $E_\perp\ll T$ was studied). }
\end{figure}

The in-plane photoelectric (IPPE) effect has been recently discovered at
terahertz (THz) frequencies in a semiconductor GaAs/Al$_x$Ga$_{1-x}$As heterostructure with a two-dimensional (2D) electron gas \cite{Michailow22}. The structure consisted in a narrow (width $W$)  2D channel, located at the interface between GaAs and Al$_x$Ga$_{1-x}$As at a depth $d$ under the surface, Figure \ref{fig:physics}(a). The channel was supplied by two, source and drain, contacts and was covered by two, left and right, metallic gates which had the shape of a bow-tie antenna, Figure \ref{fig:physics}(b). The gap between the gates had the width $b$ which was much smaller than the mean free path of 2D electrons $l_{\rm mfp}$, $b\ll l_{\rm mfp}$. If different gate voltages $U_L$ and $U_R$ were applied to the left and right gates, normally incident (along the $z$-axis) THz radiation generated a direct photocurrent in the lateral ($x$-) direction. 

The physics of the IPPE effect has been explained in  Ref. \cite{Michailow22}. If no voltages are applied to the left and right gates, $U_{L}=U_{R}=0$, the density of 2D electrons in the channel is uniform, and the chemical potential $\mu_0$, i.e. the distance between the global Fermi energy $E_F$ and the conduction band bottom energy $V_0$, does not depend on the coordinate $x$, Figure \ref{fig:physics}(c). Irradiation of the structure with electromagnetic waves would not lead to a photocurrent in the lateral direction in this situation. Applying different voltages $U_{L}\neq U_{R}$ to the left and right gates, Figure \ref{fig:physics}(d), creates a potential step $V(x)$ of a height $V_R-V_L$ for 2D electrons moving in the horizontal direction from source to drain (or visa versa); here $V_L$ and $V_R$ are the conduction band bottom energies under the left and right gates. The densities of 2D electrons under the gates are now different, and the local chemical potential becomes a function of the coordinate $x$, taking the values $\mu_{L}=\mu_0-V_L$ and $\mu_{R}=\mu_0-V_R$ under the left and right gates, Figure \ref{fig:physics}(d). Irradiation of the structure with normally incident THz waves with frequency $\omega$ now leads to a photocurrent in the $x$-direction: 2D electrons absorb quanta of THz radiation, focused by the antenna in the gap between the gates, and jump on the potential step, creating an electron flow from the region of high to the region of low electron density, Figure \ref{fig:physics}(e). This \textit{in-plane} PE effect has a number of significant advantages over the conventional \cite{Lenard1902,Einstein1905,Tamm1931,Mitchell1934} PE effect. In particular, it can be used at normal incidence of radiation, the height of the potential step can be tuned by applying the gate voltages, and the effect can be observed (and is maximal) at negative values of the ``work function'' $\phi=V_R-E_F$, Figure \ref{fig:physics}(e), see discussions in Ref. \cite{Michailow22,Mikhailov22} for more details. The corresponding device, which was called a photoelectric tunable-step (PETS) detector \cite{Michailow22}, can be used for efficient detection of THz radiation. A comprehensive analytical theory of the IPPE effect, which quantitatively explained results of the experiment \cite{Michailow22}, has been developed in Ref. \cite{Mikhailov22}. 

As was shown in Ref. \cite{Mikhailov22}, in general, the photocurrent generated in the PETS detector depends on several energy scales: (a) the chemical potentials $\mu_L$ and $\mu_R$, (b) the photon energy $\hbar\omega$, (c) the thermal energy $T$, and (d) the energy $E_\perp$ related to the quantization of the electron motion in the transverse ($y$-) direction. In Ref. \cite{Mikhailov22} we assumed that the confining potential in the $y$-direction, $V_\perp(y)$, has the form of a rectangular well with infinitely high walls. The transverse quantization energy [the bottoms of the quasi-one-dimensional (1D) subbands $E_{\perp,n}(k_x)=E_{\perp,n}+\hbar^2k_x^2/2m$] had therefore the form
\be 
E_{\perp,n}=E_\perp n^2=E_Wn^2,\  \ n=1,2,\dots,
\label{Ey_infinitewalls}
\ee
where $E_\perp=E_W=\hbar^2\pi^2/2mW^2$, and $m$ is the electron effective mass, see Eqs. (9)-(10) in Ref. \cite{Mikhailov22}. 

Under typical experimental conditions the chemical potentials and the photon energy (at THz frequencies) are on the tens of meV order. The thermal energy at $T\lesssim 10$ K is at least one order of magnitude smaller ($\lesssim 1$ meV). The energy $E_W$ was on the $\mu$eV scale in Ref. \cite{Michailow22} ($E_W\approx 5.6$ $\mu$eV at $W\approx 1$ $\mu$m). Therefore, in the previous theoretical paper \cite{Mikhailov22} we have investigated the case $E_W\ll T\ll\mu_0$, with $T\to 0$ and $E_W\to 0$, corresponding to the conditions of the experiment \cite{Michailow22}. 

Physically, the condition $E_\perp\ll T\ll\mu_0$ means that the transverse quantization of the electron motion in the 2D channel is not essential: the number $N_{\rm 1D}$ of occupied quasi-one-dimensional subbands is substantially larger than 1, and the distance between the energy bands is smaller than temperature. In this paper, we analyze another limit $T\ll E_\perp$, which can be realized, for example, in the channels of width $W\lesssim 400$ nm at temperatures $T\lesssim 0.4$ K. In this case temperature is smaller than the inter-subband distance, and the system becomes (quasi-)one-dimensional. The conductance of such narrow channels is quantized at low temperatures in units $e^2/\pi\hbar$, Refs. \cite{Wees88,Wharam88}. One of the goals of this paper is to investigate, whether and how the one-dimensional quantization of the electron spectrum in such narrow channels influences their photoresponse.

Before starting to present our theory, one remark should be made. Modeling the transverse confinement $V_\perp(y)$ in the form of a rectangular well is reasonable for relatively wide channels ($W\gtrsim 1$ $\mu$m) used in Ref. \cite{Michailow22}. In the narrower channels that we are going to study in this paper, a parabolic confinement $V_\perp(y)\propto y^2$ model is more appropriate; this is seen for example from results of the experiments \cite{Wees88,Wharam88}. In this paper, we will therefore analyze the photoresponse of narrow 2D channels with the transverse quantization energy, 
\be 
E_{\perp,n}=E_\perp\left(n+\frac 12\right)=\hbar\omega_0\left(n+\frac 12\right),\  \ n=0,1,\dots \label{Ey_parabolic}
\ee
where $E_\perp=\hbar\omega_0$, and $\omega_0$ is the harmonic oscillator frequency corresponding to the parabolic confinement $V_\perp(y)\propto y^2$.

\section{Theory and results\label{sec:theory}}

We will consider the system shown in Figure \ref{fig:physics} under THz irradiation with frequency $\omega$. Our goal will be to calculate the photocurrent and the quantum efficiency of such a device, assuming the transverse quantization spectrum in form (\ref{Ey_parabolic}).

\subsection{Hamiltonian}

As in Ref. \cite{Mikhailov22} we describe the motion of electrons in the narrow channel shown in Figure \ref{fig:physics}(b) by the time-dependent Schr\"odinger equation
\be 
i\hbar \frac{\p \Psi}{\p t}=\hat H_0\Psi+\hat H_1(x,t)\Psi,
\label{SE}
\ee
where the unperturbed part of the Hamiltonian is
\be 
\hat H_0=-\frac{\hbar^2}{2m}\frac{\p^2}{\p x^2}-\frac{\hbar^2}{2m}\frac{\p^2}{\p y^2}+V_0(x)+V_\perp(y),
\ee
\be 
V_0(x)=V_L+(V_R-V_L)\Theta(x),
\ee
and the perturbation $\hat H_1=V_1(x,t)$ has the form 
\be 
V_1(x,t)=\frac 12e\Delta\Phi_{\rm ac}\sign(x)\cos\omega t.
\ee
Here $\Delta\Phi_{\rm ac}$ is the amplitude of the time-dependent potential difference between the antenna wings, and $\Theta(x)$ is the Heaviside function; for further details see Ref. \cite{Mikhailov22}.

\subsection{Photocurrent}

Using the method of Ref. \cite{Mikhailov22}, we solve the problem (\ref{SE}) within the first order perturbation theory in $V_1$. A general expression for the photocurrent is similar to Eq. (57) in Ref. \cite{Mikhailov22}, with that difference that the transverse quantization energy $E_Wn^2$ is replaced by $\hbar\omega_0\left(n+1/2\right)$,
\ba 
I^{(1)} &=&-\frac{e}{\pi\hbar} \left(\frac {e\Delta\Phi_{ac}}{\hbar\omega}\right)^2
\sum_{n=0}^\infty \sum_\pm
\int_{-\infty}^\infty  dE 
F(E-\mu_0,T)\Big(1-F(E\pm \hbar\omega-\mu_0,T)\Big)
\nonumber \\ &\times&
\left[
\mathsf{T}_{\pm}^{\rightrightarrows}\left(\frac{E-\hbar\omega_0 (n+1/2)-V_L}{V_B},\frac{\hbar\omega}{V_B}\right) - \mathsf{T}_{\pm}^{\leftleftarrows}\left(\frac{E-\hbar\omega_0 (n+1/2)-V_L}{V_B},\frac{\hbar\omega}{V_B}\right)
\right].
\label{photocurrent-1D-parab}
\ea
Here $F(E,T)$ is the Fermi distribution function 
\be 
F(E,T)=\left[1+\exp\left(\frac{E}{T}\right)\right]^{-1},
\label{FermiFunction}
\ee 
$V_B=V_R-V_L=\mu_L-\mu_R$ is the height of the potential step (it is assumed that $V_B=V_R-V_L>0$), and the functions $\mathsf{T}_\pm^{\rightrightarrows}\left({\cal E}, \Omega\right)$ and $\mathsf{T}_\pm^{\leftleftarrows}\left({\cal E},\Omega\right)$ are given by the formulas (see Ref. \cite{Mikhailov22}),
\be
\mathsf{T}_\pm^{\rightrightarrows}\left({\cal E}, \Omega\right)=\Theta({\cal E})\Theta({\cal E}\pm\Omega-1)
\frac{\sqrt{{\cal E}}\sqrt{{\cal E}\pm\Omega-1}
\left|\sqrt{{\cal E}-1}+\sqrt{{\cal E}\pm\Omega}\right|^2}
{\left|\sqrt{{\cal E}}+\sqrt{{\cal E}-1} \right|^2 \left|\sqrt{{\cal E}\pm\Omega} +\sqrt{{\cal E}\pm\Omega-1}\right|^2},
\label{calT->->}
\ee
\be
\mathsf{T}_\pm^{\leftleftarrows}\left({\cal E},\Omega\right)=\Theta({\cal E}-1)\Theta({\cal E}\pm\Omega)
\frac{\sqrt{{\cal E}-1}\sqrt{{\cal E}\pm\Omega}
\left|\sqrt{{\cal E}}+\sqrt{{\cal E}\pm\Omega-1}\right|^2}
{\left|\sqrt{{\cal E}}+\sqrt{{\cal E}-1} \right|^2 \left|\sqrt{{\cal E}\pm\Omega} +\sqrt{{\cal E}\pm\Omega-1}\right|^2}.
\label{calT<-<-}
\ee
The former describes the probability that an electron moving from left to right will absorb (the subscript $+$) or emit (the subscript $-$) a THz photon at the potential step and will continue to move in the same direction (to the right). The latter has the same meaning for an electron, moving from right to left. Both functions depend on the electron (${\cal E}$) and photon ($\Omega$) energy normalized to the potential step height $V_B$.

In this paper, we investigate the case $T\ll \hbar\omega_0$. Since temperature $T$ is the smallest energy parameter in the problem, we can take the limit $T\to 0$ in (\ref{photocurrent-1D-parab}). Then the emission contribution vanishes, and the photocurrent can be presented, after some transformations, in the form
\be 
I^{(1)} =-\frac{e\omega}{\pi} \left(\frac {e\Delta\Phi_{ac}}{\hbar\omega}\right)^2 {\cal I}\left(\frac{\mu_L}{\hbar\omega_0},\frac{\mu_R}{\hbar\omega_0},\frac{\hbar\omega}{\hbar\omega_0}\right),
\label{photocurrent-1D-parab-T0}
\ee
where in contrast to Ref. \cite{Mikhailov22}, the chemical potentials are normalized not to $\hbar\omega$ but to $\hbar\omega_0$. The term $e\omega/\pi = 2e f$ in (\ref{photocurrent-1D-parab-T0}) has the dimension of current and equals $0.32$ $\mu$A at the frequency of $1$ THz. The factor $\alpha = (e\Delta\Phi_{ac}/\hbar\omega)^2$ is the perturbation theory parameter, determined by the ratio of the AC potential difference between the antenna wings to the photon energy. It should be smaller than 1 for the theory to be valid; otherwise, higher orders of the perturbation theory have to be taken into account. The dimensionless function 
\be 
{\cal I}\left(\zeta_L,\zeta_R,Z\right)=\frac {1}{Z}
\sum_{n=0}^{N_{\max}} 
\int_{\zeta_L- (n+1/2)-Z}^{\zeta_L- (n+1/2)}   dX 
\left[
\mathsf{T}_{+}^{\rightrightarrows}\left(\frac{ X}{\zeta_L-\zeta_R}, \frac{Z}{\zeta_L-\zeta_R}\right) - \mathsf{T}_{+}^{\leftleftarrows}\left(\frac{ X}{\zeta_L-\zeta_R}, \frac{Z}{\zeta_L-\zeta_R}\right)
\right]
\label{calI}
\ee
depends on dimensionless parameters $\zeta_{L,R}=\mu_{L,R}/\hbar\omega_0$ and $Z=\omega/\omega_0$, i.e., on the left and right chemical potentials $\mu_L$, $\mu_R$, and the photon energy $\hbar\omega$, normalized to $\hbar\omega_0$. The integer number $N_{\max}$ in (\ref{calI}) is different for the $^{\rightrightarrows}$ and $^{\leftleftarrows}$ terms in the integrand,
\be 
N_{\max}^{\rightrightarrows}=
\left\{ 
\begin{array}{cl}
\left\lfloor \zeta_R+Z- 1/2\right\rfloor, & \ \ \textrm{if }Z<(\zeta_L-\zeta_R) \\
\left\lfloor \zeta_L- 1/2\right\rfloor, & \ \ \textrm{if }Z>(\zeta_L-\zeta_R)  \\
\end{array}
\right. ,\ \ \ \ 
N_{\max}^{\leftleftarrows}=\left\lfloor \zeta_R-1/2\right\rfloor;
\ee
$\lfloor x\rfloor$ is the floor function.

\begin{figure}
\includegraphics[width=0.49\textwidth]{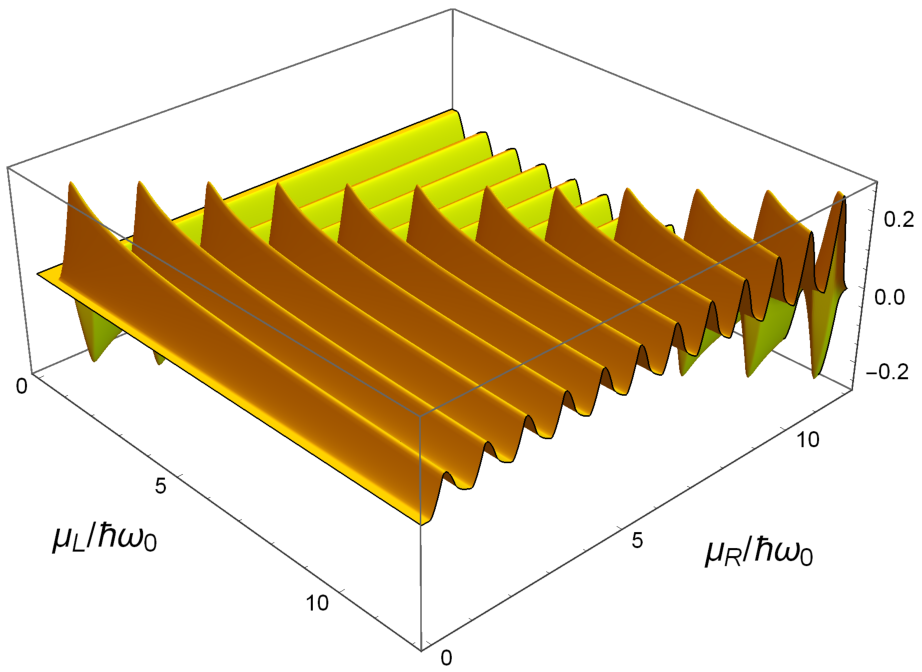}
\includegraphics[width=0.49\textwidth]{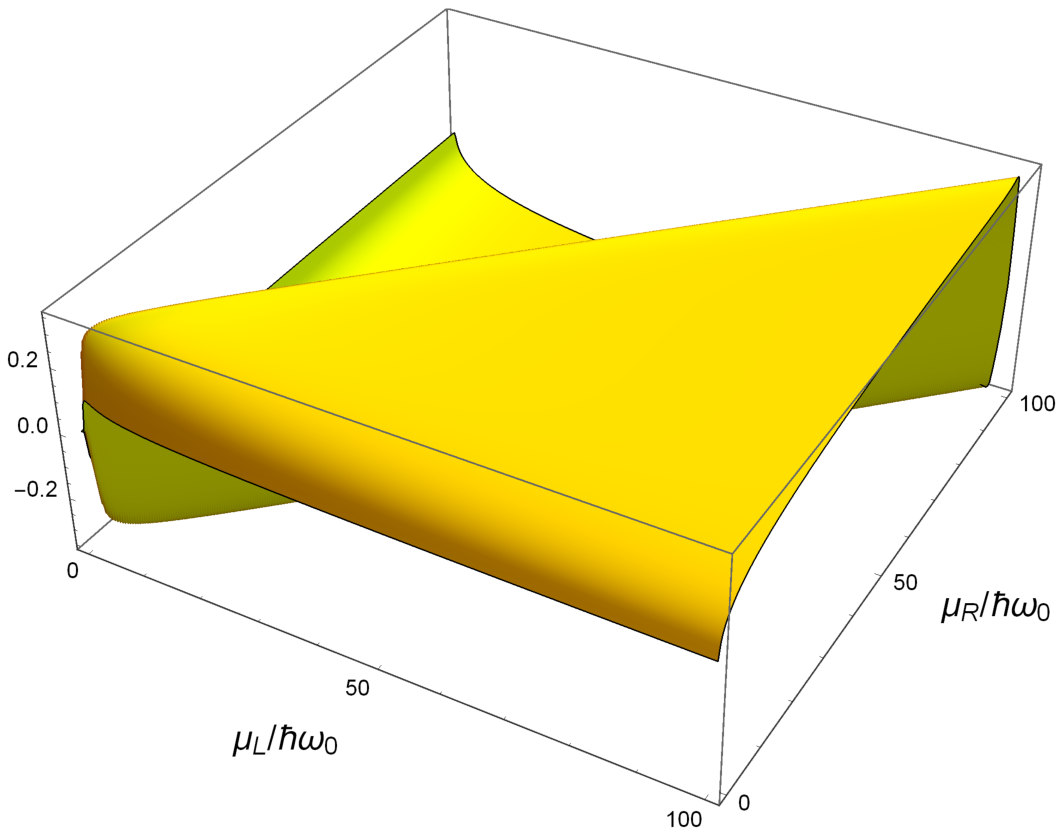}
\caption{\label{fig:Iph3D} The function ${\cal I}$, defined by Eq. (\ref{calI}), in dependence on the normalized left $\mu_L/\hbar\omega_0$ and right $\mu_R/\hbar\omega_0$ chemical potentials, for $\omega/\omega_0=0.3$ (left panel) and $\omega/\omega_0=1.0$ (right panel). The photocurrent oscillations in (a) arise due to the sequential opening of one-dimensional channels.
}
\end{figure}

\begin{figure}
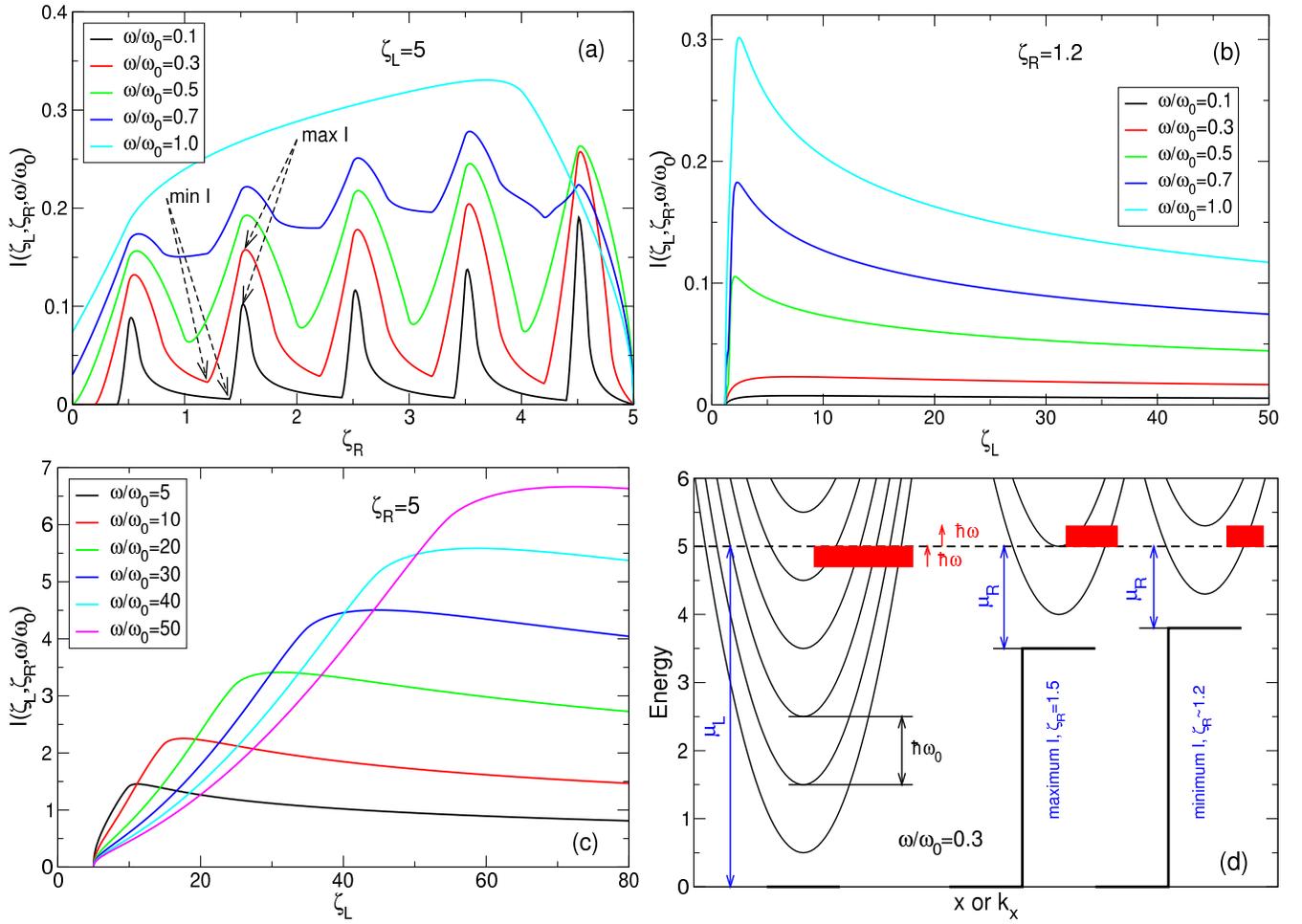

\includegraphics[width=0.49\textwidth]{Iph_MuL5-a.eps}
\includegraphics[width=0.49\textwidth]{Iph_MuR1.2-b.eps}\\
\includegraphics[width=0.49\textwidth]{Iph_MuR5-c.eps}
\includegraphics[width=0.49\textwidth]{BildParabolas.eps}
\caption{\label{fig:Iph2D} (a),(b),(c) The function ${\cal I}$, defined by Eq. (\ref{calI}), in dependence on $\zeta_R=\mu_R/\hbar\omega_0$, for (a) $\zeta_R\le \zeta_L=5$, and (b),(c) $\zeta_R\ge \zeta_L=5$, at different values of $\omega/\omega_0$: (a),(b) $\omega/\omega_0\le 1$, (c) $\omega/\omega_0\gg 1$. (d) Illustration of electron transitions for the case $\mu_L=5\hbar\omega_0$ and $\omega/\omega_0=0.3$, for the local maximum ($\mu_L=1.5\hbar\omega_0$) and local minimum ($\mu_L=1.5\hbar\omega_0-\hbar\omega\approx 1.2\hbar\omega_0$) of the photocurrent. The horizontal dashed line is the Fermi level. The local minima and maxima of ${\cal I}$ are indicated by dashed arrows in (a), for $\omega/\omega_0=0.1$ and $\omega/\omega_0=0.3$.
}
\end{figure}

The function ${\cal I}$ consists of a finite sum of one-dimensional integrals and can be easily numerically calculated. Figure \ref{fig:Iph3D} shows three-dimensional plots of ${\cal I}$ as a function of $\zeta_L$ and $\zeta_R$, for two different values of the parameter $\omega/\omega_0$, $0.3$ (left panel) and $1.0$ (right panel). One sees that at $\omega/\omega_0\ll 1$ (left panel) the dependence ${\cal I}(\zeta_L,\zeta_R)$ has many resonances which appear due to the 1D quantization of the electron spectrum. If $\omega/\omega_0= 1$ (right panel), the 1D quantization does not manifest itself anymore in the function ${\cal I}\left(\zeta_L,\zeta_R\right)$, which is now smooth. In Figure \ref{fig:Iph2D} we analyze the behavior of ${\cal I}$ in more detail. Figure \ref{fig:Iph2D}(a) shows the function ${\cal I}$ in dependence on the right normalized chemical potential $\zeta_R$, in the interval $\zeta_R\le\zeta_L=5$, and for several $\omega/\omega_0$'s varying from $0.1$ up to $1.0$. At $\omega/\omega_0=0.1$ the curve ${\cal I}(\zeta_R)$ has several sharp peaks, with maxima at $\mu_R\approx \hbar\omega_0(n+1/2)$ and minima at $\mu_R\approx \hbar\omega_0(n+1/2)-\hbar\omega$, where  $n$ is an integer. The width of the resonances is thus determined by $\hbar\omega$; their shape reminds the density of states in a one-dimensional electron gas. When the frequency increases, the maxima of ${\cal I}$ corresponding to different $n$ start to overlap, and at $\omega/\omega_0=1$ the function ${\cal I}$ acquires a structureless smooth form with a wide maximum.

The physical reasons of such a behavior of the function ${\cal I}$ are explained in Figure \ref{fig:Iph2D}(d). In the left part of this panel we show five quasi-1D energy subbands $E_n(k_x)$ of electrons at $x<0$. They intersect the Fermi level $E_F$ (horizontal dashed line) at the energy $5\hbar\omega_0$ corresponding to $\zeta_L=5$. The $k_x$-states covered by a red rectangle correspond to electrons with energies from $E_F-\hbar\omega$ to $E_F$ which run to the right and are able to jump to the potential step after absorption of a photon ($\hbar\omega=0.3\hbar\omega_0$ on the Figure). In the right part of the panel we show the quasi-1D electron energy subbands $E_n(k_x)$ at the potential step at $x>0$. Here two cases are considered, with $\mu_R=(1+1/2)\hbar\omega_0=1.5\hbar\omega_0$ which corresponds to the maximum of the photocurrent, and with $\mu_R=(1+1/2)\hbar\omega_0-\hbar\omega=1.2\hbar\omega_0$ which corresponds to the minimum of the photocurrent. One sees that in the former (maximum ${\cal I}$) case, the photoexcited electrons are able to jump to the states of the first and the second subbands (covered by a red rectangle). In the latter case they jump only to the states of the first subband and do not have enough energy to jump to the states of the second subband. Therefore this leads to a minimum of ${\cal I}$. It is clear that when the parameter $\omega/\omega_0$ grows and approaches 1, the spectrum of allowed final states at $x>0$ overlap and the curve ${\cal I}(\zeta_R)$ becomes smooth, Figure \ref{fig:Iph2D}(a).

Figure \ref{fig:Iph2D}(b) shows the behavior of the function ${\cal I}$ in dependence on $\zeta_L$, in the interval $\zeta_L\ge\zeta_R=1.2$ for several values of $\omega/\omega_0\le 1$. This time the right chemical potential is fixed, whereas the left chemical potential grows. Now, the dependence of ${\cal I}(\zeta_L)$ does not oscillate, the function ${\cal I}(\zeta_L)$ has a maximum at $\mu_L-\mu_R\simeq \hbar\omega$, and this maximum grows with $\omega/\omega_0$. The reason of these features is also seen in Figure \ref{fig:Iph2D}(d): the number of occupied subbands in the left part of the panel, and hence the number of initial electron states which may contribute to the flow of photoexcited electrons, grows with $\omega/\omega_0$. This leads to the increase in photocurrent. The maximum of ${\cal I}(\zeta_L)$ is seen at $\mu_L-\mu_R\simeq \hbar\omega$ because, in this case, all states under the left chemical potential participate in the photocurrent generation. In the opposite case, when $\mu_L-\mu_R$ substantially exceeds $\hbar\omega$, an essential part of electrons on the left side of the device cannot contribute to ${\cal I}$.

Similar features are seen in the dependence of ${\cal I}(\zeta_L)$ at larger values of $\omega/\omega_0\gg 1$, see Figure \ref{fig:Iph2D}(c). The curves ${\cal I}(\zeta_L)$ become smoother, their maxima ${\cal I}_{\max}$ (as a function of $\zeta_L$) shift to larger values of $\zeta_L$, roughly corresponding to $\mu_L\simeq \mu_R+\hbar\omega$, and their absolute value increases. 

\begin{figure}
\includegraphics[width=0.49\textwidth]{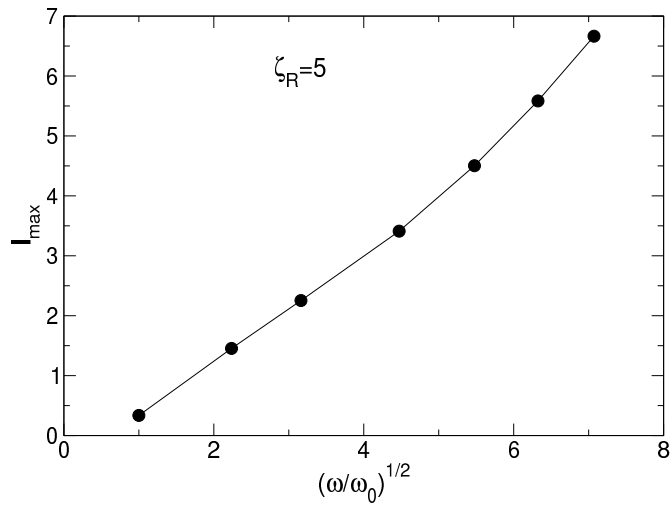}
\caption{\label{fig:maximaIph} Maxima ${\cal I}_{\max}$ of the function ${\cal I}(\zeta_L, \zeta_R)$ vs. $(\omega/\omega_0)^{1/2}$ for $\zeta_R=5$. ${\cal I}_{\max}=0.3355$ at $\omega/\omega_0=1$. 
}
\end{figure}

Quantitatively, the maxima ${\cal I}_{\max}$ of the curves, shown in Figures \ref{fig:Iph2D}(b) and (c), grow with $\omega/\omega_0$ approximately as $\sqrt{\omega/\omega_0}$, see Figure \ref{fig:maximaIph}. The regime ($\zeta_L\gg 1$, $\omega/\omega_0\gg 1$) corresponds to the situation considered in Ref. \cite{Mikhailov22}: In this case, the transverse quantization energy becomes smaller than both the photon energy and the chemical potential, and the number of occupied 1D subbands becomes much larger than 1. Therefore it makes sense to compare the absolute values of the photocurrent maxima calculated here and in Ref. \cite{Mikhailov22} for the parabolic and rectangular well confinements, respectively. As seen from the definitions of the photocurrent $I^{(1)}$ in this paper and in Ref. \cite{Mikhailov22}, the function ${\cal I}$ should be compared with $\sqrt{\hbar\omega/E_W}{\cal J}$ in Ref. \cite{Mikhailov22}. The maximum photocurrents are quantitatively similar in both works. Indeed, in \cite{Mikhailov22} ${\cal J}_{\max}$ was found to be about $0.22$, so that the maximum photocurrent is given by a small number ($\sim 0.22$) times the square root of the photon energy divided by the energy of the transverse quantization, $\sqrt{\hbar\omega/E_\perp}$, where $E_\perp=E_W$. In the present paper, it is also a small number (e.g., $\sim 0.335$  at $\zeta_R=5$, Figure \ref{fig:maximaIph}) times $\sqrt{\hbar\omega/E_\perp}$, with $E_\perp=\hbar\omega_0$ in the parabolic confinement case. 

It is noticeable that the photocurrent maxima in Figures \ref{fig:Iph3D}(a) and \ref{fig:Iph2D}(a) (at $\omega/\omega_0\lesssim 1$) are located equidistantly on the $\zeta_R$ axis. This is a direct consequence of the equidistant spectrum of electrons (\ref{Ey_parabolic}) in the parabolic confinement potential $V_\perp(y)$. If electrons were moving in a rectangular potential well with infinitely high walls \cite{Mikhailov22}, then the energy $E_{\perp,n}$ would be proportional to $n^2$, and the positions of the photocurrent maxima on the $\zeta_R$ axis would follow the $n^2$ sequence. Thus, analysis of the IPPE effect in quasi-one-dimensional electron systems makes it possible, in principle, to draw a conclusion about the spectrum of one-dimensional quantization $E_{\perp,n}$ and the shape of the confining potential $V_\perp(y)$ in the system.

\subsection{Quantum efficiency}

Another quantity which is interesting to analyze is the quantum efficiency $\eta(\zeta_L,\zeta_R,\omega/\omega_0)$. We define it in the same way as in Ref. \cite{Mikhailov22} as the ratio of the number of  electrons which absorbed a photon and contributed to the photocurrent to the total number of electrons, which absorbed a photon. The quantum efficiency can be symbolically written as
\be 
\eta(\zeta_L,\zeta_R,Z)=\frac{{\cal W}\left[
\mathsf{T}_{+}^{\rightrightarrows} - \mathsf{T}_{+}^{\leftleftarrows}
\right]}{{\cal W}\left[
\mathsf{T}_{+}^{\rightrightarrows} + \mathsf{T}_{+}^{\leftleftarrows} + \mathsf{R}_{+}^{\leftrightarrows} + \mathsf{R}_{+}^{\rightleftarrows}
\right]},
\ee
where the function in the nominator is the photocurrent (\ref{calI}), whereas the function in the denominator is obtained from (\ref{calI}) by replacing the difference $\mathsf{T}_{+}^{\rightrightarrows} - \mathsf{T}_{+}^{\leftleftarrows}$ by the sum $\mathsf{T}_{+}^{\rightrightarrows} + \mathsf{T}_{+}^{\leftleftarrows} + \mathsf{R}_{+}^{\leftrightarrows} + \mathsf{R}_{+}^{\rightleftarrows}$ in the integrand of Eq. (\ref{calI}). 

\begin{figure}
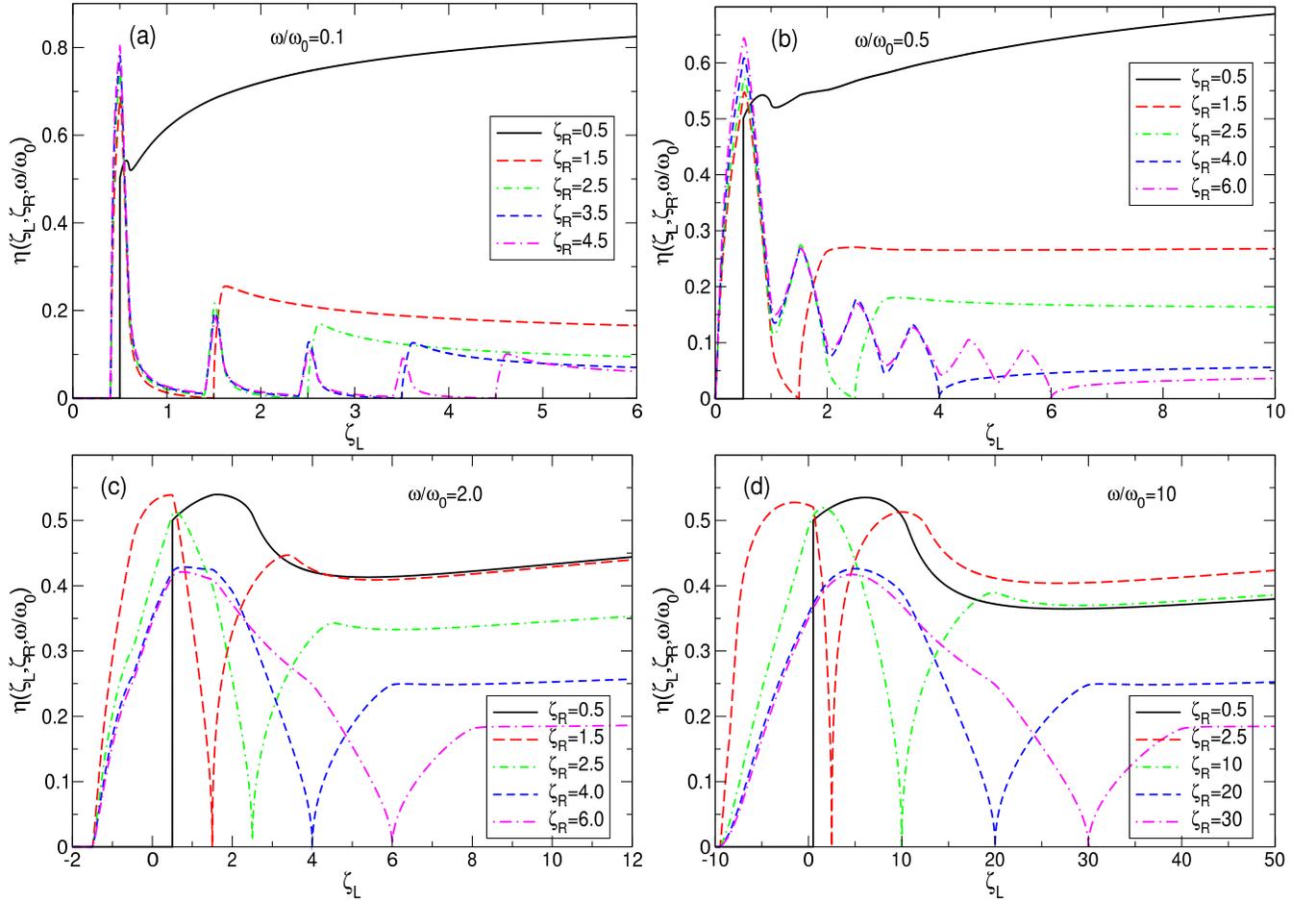

\includegraphics[width=0.49\textwidth]{QE_W0.1.eps}
\includegraphics[width=0.49\textwidth]{QE_W0.5.eps}\\
\includegraphics[width=0.49\textwidth]{QE_W2.0.eps}
\includegraphics[width=0.49\textwidth]{QE_W10.eps}
\caption{\label{fig:QE} The quantum efficiency $\eta(\zeta_L,\zeta_R,Z)$, as a function of $\zeta_L$ at a few values of $\zeta_R$, at (a) $Z\equiv \omega/\omega_0=0.1$, (b) $Z=0.5$, (c) $Z=2$, and (d) $Z=10$. 
}
\end{figure}

Figure \ref{fig:QE} shows the quantum efficiency, as a function of one of the chemical potential ($\zeta_L$), at fixed values of the other chemical potential $\zeta_R$. Different panels correspond to different values of $\omega/\omega_0$. Let us consider first the case $\zeta_R=0.5$, described by black solid curves on all panels of Figure \ref{fig:QE}. It corresponds to the case, when the bottom of the lowest energy band at $x>0$ touches the Fermi level, i.e., there are no electrons in the right half of the device. If $\zeta_L<0.5$, the quantum efficiency, as well as the photocurrent vanish at all values of $\omega/\omega_0$, since in this case there are no electrons in the left half of the device too. Now, if $\zeta_L$ is larger than $\zeta_R$ and tends to it from above, $\zeta_L\to\zeta_R=0.5$, the quantum efficiency tends to 0.5 for all values of $\omega/\omega_0$. This limit, $0<\zeta_L-\zeta_R\ll 1$, corresponds to a very small potential step $V_B=\mu_L-\mu_R\ll\hbar\omega_0$. The parameter $\Omega$ in Eqs. (\ref{calT->->})--(\ref{calT<-<-}) becomes very large, whereas the value of ${\cal E}$ is restricted, $0<{\cal E}<1$. Under these conditions, the probabilities $\mathsf{T}_\pm^{\leftleftarrows}$ and $\mathsf{R}_\pm^{\rightleftarrows}$ equal zero (there are no electrons at $x>0$), whereas the probabilities $\mathsf{T}_\pm^{\rightrightarrows}$ and $\mathsf{R}_\pm^{\leftrightarrows}$ equals each other, $\mathsf{T}_\pm^{\rightrightarrows}=\mathsf{R}_\pm^{\leftrightarrows}\approx \sqrt{{\cal E}\Omega^3}/4$. (The formulas for $\mathsf{R}_\pm^{\leftrightarrows}({\cal E},\Omega)$ and $\mathsf{R}_\pm^{\rightleftarrows}({\cal E},\Omega)$ can be found in Ref. \cite{Mikhailov22}). Thus, electrons approaching a small potential step from the left, have equal probabilities to be reflected back or to keep moving to the right after absorption of a photon. The quantum efficiency then equals $1/2$. 

Then, when $\zeta_L$ grows, the quantum efficiency first grows with $\zeta_L-\zeta_R$, then has a local minimum at $\zeta_L-\zeta_R\simeq \hbar\omega$ and then grows again. The asymptotic values of $\eta$ at large $\zeta_L$ are larger at small values of $\omega/\omega_0$: as seen from Figure \ref{fig:QE}(a), at $\omega/\omega_0=0.1$ the quantum efficiency exceeds 80\% already at $\zeta_L\simeq 6$. At larger $\omega/\omega_0$ the asymptotic values of $\eta$ are smaller but still on the order of $40-50$\% for parameters of Figure \ref{fig:QE}.

Now, let us consider the cases of larger values of $\zeta_R$ ($>1/2$), when there are electrons in the right half of the device. At small frequencies, $\omega/\omega_0<1$, Figures \ref{fig:QE}(a),(b), the 1D quantization of the electron spectrum manifests itself also in the quantum efficiency. At small values of $\zeta_L<\zeta_R$, $\eta$ has a sharp maximum at $\zeta_L\simeq 0.5$. When $\zeta_L$ grows, $\eta$ oscillates, with sharp maxima of the width $\simeq \omega/\omega_0$ at $\zeta_L=n+1/2$; however, when $\zeta_L$ reaches $\zeta_R$, the function $\eta$ becomes zero and then becomes a smooth function which slowly decreases with $\zeta_L$. When $\omega/\omega_0$ becomes larger than 1, Figures \ref{fig:QE}(c),(d), the maxima of $\eta$ overlap and the curves $\eta(\zeta_L)$ acquire a simple form with zeros at $\zeta_L=\hbar(\omega_0/2-\omega)$ and at $\zeta_L=\zeta_R$. At $\zeta_L>\zeta_R$ the curves $\eta(\zeta_L)$ describe slowly varying with $\zeta_L$ functions. Quantitatively the quantum efficiency can be as large as $\sim 0.5$ at $\omega/\omega_0>1$ and larger (up to $0.8-0.9$ in Figure \ref{fig:QE}) at $\omega/\omega_0<1$. In the in-plane photoelectric effect the transformation of photons to the photocurrent is, thus, extremely efficient.

\section{Summary}

We have investigated the in-plane photoelectric effect in narrow, quasi-one-dimensional electron channels at low temperatures, when the transverse quantization energy $E_\perp$ exceeds the thermal energy $T$. The confining potential in the perpendicular direction was assumed to be parabolic, with the characteristic harmonic-oscillator frequency $\omega_0$ and $E_\perp=\hbar\omega_0$. We have shown that, if the radiation frequency $\omega$ is smaller than $\omega_0$, the photoresponse oscillates as a function of gate voltages, due to the one-dimensional quantization of the electron motion in the channel. The positions of the oscillating photocurent maxima carry information about the spectrum of one-dimensional quantization and the shape of the confining potential $V_\perp(y)$ in the system. The absolute value of the photocurrent is proportional to the square root of the number of occupied quasi-1D electronic subbands. The quantum efficiency can be as large as $50$\%$-90$\% at different parameters of the device. The theory is applicable to any semiconductor systems with 2D electron gases, including III-V quantum well structures, silicon-based field effect transistors, as well as novel 2D layered semiconductor materials.

\begin{acknowledgments}
The work was supported by the European Union's Horizon 2020 Research and Innovation Program Graphene Core 3 under Grant Agreement No. 881603. I thank Wladislaw Michailow for useful discussions of experimental aspects of the IPPE measurements.
\end{acknowledgments}

\bibliography{../../../../../../../../BIB-FILES/mikhailov,../../../../../../../../BIB-FILES/graphene,../../../../../../../../BIB-FILES/lowD,../../../../../../../../BIB-FILES/thz} 

\begin{thebibliography}{8}%
\makeatletter
\providecommand \@ifxundefined [1]{%
 \@ifx{#1\undefined}
}%
\providecommand \@ifnum [1]{%
 \ifnum #1\expandafter \@firstoftwo
 \else \expandafter \@secondoftwo
 \fi
}%
\providecommand \@ifx [1]{%
 \ifx #1\expandafter \@firstoftwo
 \else \expandafter \@secondoftwo
 \fi
}%
\providecommand \natexlab [1]{#1}%
\providecommand \enquote  [1]{``#1''}%
\providecommand \bibnamefont  [1]{#1}%
\providecommand \bibfnamefont [1]{#1}%
\providecommand \citenamefont [1]{#1}%
\providecommand \href@noop [0]{\@secondoftwo}%
\providecommand \href [0]{\begingroup \@sanitize@url \@href}%
\providecommand \@href[1]{\@@startlink{#1}\@@href}%
\providecommand \@@href[1]{\endgroup#1\@@endlink}%
\providecommand \@sanitize@url [0]{\catcode `\\12\catcode `\$12\catcode
  `\&12\catcode `\#12\catcode `\^12\catcode `\_12\catcode `\%12\relax}%
\providecommand \@@startlink[1]{}%
\providecommand \@@endlink[0]{}%
\providecommand \url  [0]{\begingroup\@sanitize@url \@url }%
\providecommand \@url [1]{\endgroup\@href {#1}{\urlprefix }}%
\providecommand \urlprefix  [0]{URL }%
\providecommand \Eprint [0]{\href }%
\providecommand \doibase [0]{https://doi.org/}%
\providecommand \selectlanguage [0]{\@gobble}%
\providecommand \bibinfo  [0]{\@secondoftwo}%
\providecommand \bibfield  [0]{\@secondoftwo}%
\providecommand \translation [1]{[#1]}%
\providecommand \BibitemOpen [0]{}%
\providecommand \bibitemStop [0]{}%
\providecommand \bibitemNoStop [0]{.\EOS\space}%
\providecommand \EOS [0]{\spacefactor3000\relax}%
\providecommand \BibitemShut  [1]{\csname bibitem#1\endcsname}%
\let\auto@bib@innerbib\@empty
\bibitem [{\citenamefont {Michailow}\ \emph {et~al.}(2022)\citenamefont
  {Michailow}, \citenamefont {Spencer}, \citenamefont {Almond}, \citenamefont
  {Kindness}, \citenamefont {Wallis}, \citenamefont {Mitchell}, \citenamefont
  {Degl'Innocenti}, \citenamefont {Mikhailov}, \citenamefont {Beere},\ and\
  \citenamefont {Ritchie}}]{Michailow22}%
  \BibitemOpen
  \bibfield  {author} {\bibinfo {author} {\bibfnamefont {W.}~\bibnamefont
  {Michailow}}, \bibinfo {author} {\bibfnamefont {P.}~\bibnamefont {Spencer}},
  \bibinfo {author} {\bibfnamefont {N.~W.}\ \bibnamefont {Almond}}, \bibinfo
  {author} {\bibfnamefont {S.~J.}\ \bibnamefont {Kindness}}, \bibinfo {author}
  {\bibfnamefont {R.}~\bibnamefont {Wallis}}, \bibinfo {author} {\bibfnamefont
  {T.~A.}\ \bibnamefont {Mitchell}}, \bibinfo {author} {\bibfnamefont
  {R.}~\bibnamefont {Degl'Innocenti}}, \bibinfo {author} {\bibfnamefont
  {S.~A.}\ \bibnamefont {Mikhailov}}, \bibinfo {author} {\bibfnamefont {H.~E.}\
  \bibnamefont {Beere}},\ and\ \bibinfo {author} {\bibfnamefont {D.~A.}\
  \bibnamefont {Ritchie}},\ }\bibfield  {title} {\bibinfo {title} {An in-plane
  photoelectric effect in two-dimensional electron systems for terahertz
  detection},\ }\bibfield  {journal} {\bibinfo  {journal} {Sci. Adv.}\ }\textbf
  {\bibinfo {volume} {8}},\ \href {https://doi.org/10.1126/sciadv.abi8398}
  {10.1126/sciadv.abi8398} (\bibinfo {year} {2022})\BibitemShut {NoStop}%
\bibitem [{\citenamefont {Lenard}(1902)}]{Lenard1902}%
  \BibitemOpen
  \bibfield  {author} {\bibinfo {author} {\bibfnamefont {P.}~\bibnamefont
  {Lenard}},\ }\bibfield  {title} {\bibinfo {title} {{\"U}ber die
  lichtelektrische {W}irkung},\ }\href@noop {} {\bibfield  {journal} {\bibinfo
  {journal} {Annalen der Physik}\ }\textbf {\bibinfo {volume} {313}},\ \bibinfo
  {pages} {149} (\bibinfo {year} {1902})}\BibitemShut {NoStop}%
\bibitem [{\citenamefont {Einstein}(1905)}]{Einstein1905}%
  \BibitemOpen
  \bibfield  {author} {\bibinfo {author} {\bibfnamefont {A.}~\bibnamefont
  {Einstein}},\ }\bibfield  {title} {\bibinfo {title} {{\"U}ber einen die
  {E}rzeugung und {V}erwandlung des {L}ichtes betreffenden heuristischen
  {G}esichtspunkt},\ }\href@noop {} {\bibfield  {journal} {\bibinfo  {journal}
  {Annalen der Physik}\ }\textbf {\bibinfo {volume} {322}},\ \bibinfo {pages}
  {132} (\bibinfo {year} {1905})}\BibitemShut {NoStop}%
\bibitem [{\citenamefont {Tamm}\ and\ \citenamefont
  {Schubin}(1931)}]{Tamm1931}%
  \BibitemOpen
  \bibfield  {author} {\bibinfo {author} {\bibfnamefont {I.}~\bibnamefont
  {Tamm}}\ and\ \bibinfo {author} {\bibfnamefont {S.}~\bibnamefont {Schubin}},\
  }\bibfield  {title} {\bibinfo {title} {Zur {T}heorie des {P}hotoeffektes an
  {M}etallen},\ }\href@noop {} {\bibfield  {journal} {\bibinfo  {journal}
  {Zeitschrift f\"ur Physik}\ }\textbf {\bibinfo {volume} {68}},\ \bibinfo
  {pages} {97} (\bibinfo {year} {1931})}\BibitemShut {NoStop}%
\bibitem [{\citenamefont {Mitchell}(1934)}]{Mitchell1934}%
  \BibitemOpen
  \bibfield  {author} {\bibinfo {author} {\bibfnamefont {K.}~\bibnamefont
  {Mitchell}},\ }\bibfield  {title} {\bibinfo {title} {The theory of the
  surface photoelectric effect in metals -- {I}},\ }\href@noop {} {\bibfield
  {journal} {\bibinfo  {journal} {Proc. of the Poyal Society A}\ }\textbf
  {\bibinfo {volume} {146}},\ \bibinfo {pages} {442} (\bibinfo {year}
  {1934})}\BibitemShut {NoStop}%
\bibitem [{\citenamefont {Mikhailov}\ \emph {et~al.}(2022)\citenamefont
  {Mikhailov}, \citenamefont {Michailow}, \citenamefont {Beere},\ and\
  \citenamefont {Ritchie}}]{Mikhailov22}%
  \BibitemOpen
  \bibfield  {author} {\bibinfo {author} {\bibfnamefont {S.~A.}\ \bibnamefont
  {Mikhailov}}, \bibinfo {author} {\bibfnamefont {W.}~\bibnamefont
  {Michailow}}, \bibinfo {author} {\bibfnamefont {H.~E.}\ \bibnamefont
  {Beere}},\ and\ \bibinfo {author} {\bibfnamefont {D.~A.}\ \bibnamefont
  {Ritchie}},\ }\bibfield  {title} {\bibinfo {title} {Theory of the in-plane
  photoelectric effect in two-dimensional electron systems},\ }\href
  {https://doi.org/10.1103/PhysRevB.106.075411} {\bibfield  {journal} {\bibinfo
   {journal} {Phys. Rev. B}\ }\textbf {\bibinfo {volume} {106}},\ \bibinfo
  {pages} {075411} (\bibinfo {year} {2022})}\BibitemShut {NoStop}%
\bibitem [{\citenamefont {van Wees}\ \emph {et~al.}(1988)\citenamefont {van
  Wees}, \citenamefont {van Houten}, \citenamefont {Beenakker}, \citenamefont
  {Williamson}, \citenamefont {Kouwenhoven}, \citenamefont {van~der Marel},\
  and\ \citenamefont {Foxon}}]{Wees88}%
  \BibitemOpen
  \bibfield  {author} {\bibinfo {author} {\bibfnamefont {B.~J.}\ \bibnamefont
  {van Wees}}, \bibinfo {author} {\bibfnamefont {H.}~\bibnamefont {van
  Houten}}, \bibinfo {author} {\bibfnamefont {C.~W.~J.}\ \bibnamefont
  {Beenakker}}, \bibinfo {author} {\bibfnamefont {J.~G.}\ \bibnamefont
  {Williamson}}, \bibinfo {author} {\bibfnamefont {L.~P.}\ \bibnamefont
  {Kouwenhoven}}, \bibinfo {author} {\bibfnamefont {D.}~\bibnamefont {van~der
  Marel}},\ and\ \bibinfo {author} {\bibfnamefont {C.~T.}\ \bibnamefont
  {Foxon}},\ }\bibfield  {title} {\bibinfo {title} {Quantized conductance of
  point contacts in a two-dimensional electron gas},\ }\href@noop {} {\bibfield
   {journal} {\bibinfo  {journal} {Phys. Rev. Lett.}\ }\textbf {\bibinfo
  {volume} {60}},\ \bibinfo {pages} {848} (\bibinfo {year} {1988})}\BibitemShut
  {NoStop}%
\bibitem [{\citenamefont {Wharam}\ \emph {et~al.}(1988)\citenamefont {Wharam},
  \citenamefont {Thornton}, \citenamefont {Newbury}, \citenamefont {Pepper},
  \citenamefont {Ahmed}, \citenamefont {Frost}, \citenamefont {Hasko},
  \citenamefont {Peacock}, \citenamefont {Ritchie},\ and\ \citenamefont
  {Jones}}]{Wharam88}%
  \BibitemOpen
  \bibfield  {author} {\bibinfo {author} {\bibfnamefont {D.}~\bibnamefont
  {Wharam}}, \bibinfo {author} {\bibfnamefont {T.~J.}\ \bibnamefont
  {Thornton}}, \bibinfo {author} {\bibfnamefont {R.}~\bibnamefont {Newbury}},
  \bibinfo {author} {\bibfnamefont {M.}~\bibnamefont {Pepper}}, \bibinfo
  {author} {\bibfnamefont {H.}~\bibnamefont {Ahmed}}, \bibinfo {author}
  {\bibfnamefont {J.~E.~F.}\ \bibnamefont {Frost}}, \bibinfo {author}
  {\bibfnamefont {D.~G.}\ \bibnamefont {Hasko}}, \bibinfo {author}
  {\bibfnamefont {D.~C.}\ \bibnamefont {Peacock}}, \bibinfo {author}
  {\bibfnamefont {D.~A.}\ \bibnamefont {Ritchie}},\ and\ \bibinfo {author}
  {\bibfnamefont {G.~A.~C.}\ \bibnamefont {Jones}},\ }\bibfield  {title}
  {\bibinfo {title} {One-dimensional transport and the quantisation of the
  ballistic resistance},\ }\href@noop {} {\bibfield  {journal} {\bibinfo
  {journal} {J. Phys. C: Solid State Phys.}\ }\textbf {\bibinfo {volume}
  {21}},\ \bibinfo {pages} {L209} (\bibinfo {year} {1988})}\BibitemShut
  {NoStop}%
\end{thebibliography}%
\end{document}